\documentclass{appolb}
\usepackage{epsfig}

\begin{document}

\title{Asymmetric matrices in an analysis of financial correlations}

\author{J.~Kwapie\'n$^1$, S.~Dro\.zd\.z$^{1,2}$, A.Z.~G\'orski$^1$, 
P.~O\'swi\c ecimka$^1$
\address{$^1$ Institute of Nuclear Physics, Polish Academy of Sciences, 
Krak\'ow, Poland \\
$^2$ Institute of Physics, University of Rzesz\'ow, Rzesz\'ow, Poland}
}

\maketitle

\begin{abstract}

Financial markets are highly correlated systems that reveal both the 
inter-market dependencies and the correlations among their different 
components. Standard analyzing techniques include correlation coefficients 
for pairs of signals and correlation matrices for rich multivariate data.  
In the latter case one constructs a real symmetric matrix with real 
non-negative eigenvalues describing the correlation structure of the data.  
However, if one performs a correlation-function-like analysis of 
multivariate data, when a stress is put on investigation of delayed 
dependencies among different types of signals, one can calculate an 
asymmetric correlation matrix with complex eigenspectrum. From the Random 
Matrix Theory point of view this kind of matrices is closely related to 
Ginibre Orthogonal Ensemble (GinOE). We present an example of practical 
application of such matrices in correlation analyses of empirical data. By 
introducing the time lag, we are able to identify temporal structure of 
the inter-market correlations. Our results show that the American and 
German stock markets evolve almost simultaneously without a significant 
time lag so that it is hard to find imprints of information transfer 
between these markets. There is only an extremely subtle indication that 
the German market advances the American one by a few seconds.

\end{abstract}

  
\section{Introduction}

A number of studies have shown that different financial markets reveal 
hierarchical structure~\cite{mantegna99,bonanno00,giada01,plerou02,%
onnela03,dimatteo04,kim05,mizuno05} that can be approximated by factor and 
group models (e.g.~\cite{roll80,noh00} for the stock market case). At the 
level of financial data, these structures are determined principally by 
strength of correlations in returns of different stocks, currencies or 
other assets. The most popular methods of such an analysis are based on 
the calculation of correlation matrices from multivariate time series of 
returns. The correlation matrices can then be diagonalized in order to 
obtain spectra of their eigenvalues and 
eigenvectors~\cite{laloux99,plerou02,kwapien06} or can serve as 
a source for the construction of minimal spanning 
trees~\cite{mantegna99,bonanno01,mizuno05,gorski06}. In the standard 
approach, in which the correlations between all analysed assets are taken 
into consideration, the correlation matrix is by construction symmetric 
due to the correlation coefficient invariance under a swap of signals. 
This obviously leads to a real eigenspectrum of the matrix.  Usually 
properties of the empirical correlation matrix are compared with universal 
predictions of the adequate, Wishart ensemble of random matrices and the 
identified deviations are considered as an indication of actual 
correlations among data.

In principle, however, there is no restriction imposed on the symmetry
property of a correlation matrix: it may well be antisymmetric or even
completely asymmetric, depending on which signals are used in the
calculations. For example, if there are two separate sets of signals and
the correlations are calculated only across these two sets, the resulting
matrix can no longer be symmetric and, consequently, its eigenspectrum can
be complex. However, there is still a non-zero probability that some of
the eigenvalues and eigenvectors are real. As long as a distribution of
the correlation matrix elements is close to a Gaussian, the most relevant
random matrix ensemble, against which the results should be tested, is the
Ginibre Orthogonal Ensemble (GinOE)~\cite{ginibre65}. For the financial
data characterized by fat tails of p.d.f. this assumption can also be made
provided the time series under study are sufficiently long.

At present one observes in literature a growing interest in theoretical
research on properties of real asymmetric and, more generally,
non-Hermitean random matrices. This interest is motivated by a broadening
spectrum of applications of such matrices which includes, among others,
random networks~\cite{timme04}, quantum chaos~\cite{fyodorov97}, quantum
chromodynamics~\cite{stephanov96,akemann04} and brain
research~\cite{kwapien00}. An issue which we address in this work and
which can serve as an example of application of the asymmetric correlation
matrices to empirical data can be related to a globalization of financial
markets. We investigate the cross-market correlations between returns of
stocks traded on two large but geographically distant markets: New York
Stock Exchange and Deutsche B\"orse. Our objective is to identify the
strength of the instanteous as well as the time lagged dependencies
between evolution of these two markets.

\section{Methods}

We begin with presenting a brief construction scheme of an asymmetric
correlation matrix and a short description of basic properties of GinOE.
Let us consider the two disjoint sets ${\rm X}, {\rm Y}$ each consisting 
of $N$ assets and denote by $\{x_i^{(s)}\}_{i=1,...,T}$ and
$\{y_i^{(t)}\}_{i=1,...,T}$ the time series of normalized logarithmic 
returns of assets $s \in {\rm X}$ and $t \in {\rm Y}$ ($s,t=1,...,N$).
For each set we construct an $N \times T$ data matrix ${\bf M}$ and  
the correlation matrix ${\bf C}_{\rm XY}$ according to formula
\begin{equation}
{\bf C}_{\rm XY} = {1 \over T} {\bf M}_{\rm X} {\bf M}_{\rm Y}^{\rm T}.
\label{corrmatx}
\end{equation}
Each matrix element $-1 \le C_{s,t} \le 1$ is the Pearson 
cross-correlation coefficient for assets $s$ and $t$ ($C_{s,t} \neq 
C_{t,s}$). In the next step the correlation matrix can be diagonalized by 
solving the eigenvalue problem
\begin{equation}
{\bf C}_{\rm XY} {\bf v}_k = \lambda_k {\bf v}_k, \ \ k = 1,...,N
\end{equation}
which provides us with a complete spectrum of generally complex 
eigenvalues $\lambda_k$ and pairs of conjugated eigenvectors ${\bf v}_k$.
The assumption ${\rm Y} = {\rm X}$ in Eq.(\ref{corrmatx}) leads to the 
standard definition of a symmetric correlation matrix ${\bf C}_{\rm XX}$ 
with a real eigenspectrum.

\begin{figure}
\epsfxsize 8cm
\hspace{2.0cm}
\epsffile{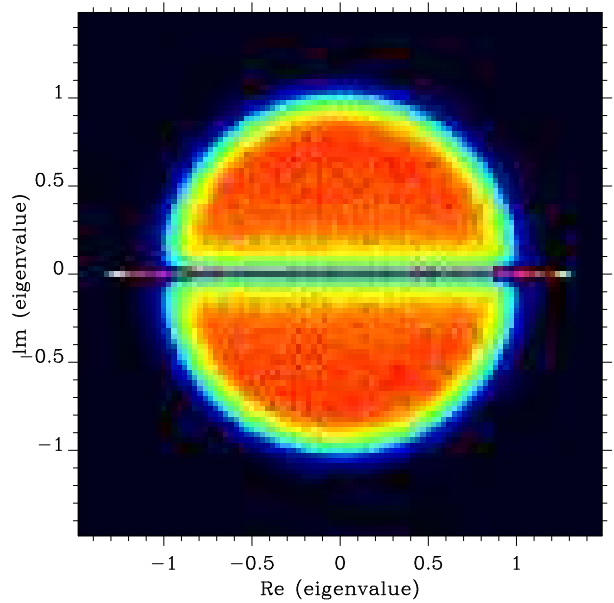}
\caption{Probability density function of complex eigenvalues of 
exemplary GinOE random matrix ($N=30$) obtained by averaging the spectra 
over 100000 individual matrix realizations. Colors range from black 
(probability density close to zero) to red (highest probability density).}
\end{figure}

Properties of the empirical correlation matrix have to be tested against a 
null hypothesis of completely random correlations characteristic for 
independent signals. Random Matrix Theory (RMT) offers some analytic
results for a corresponding ensemble of real asymmetric matrices, i.e. the 
Ginibre Orthogonal Enseble~\cite{ginibre65} defined by the Gaussian 
probability density
\begin{equation}
P_{\rm GinOE}(\mathcal{C}) = (2 \pi)^{-N^2/2} \exp [-{\rm 
Tr}(\mathcal{CC}^{\rm 
T}/2)],
\label{pdfginoe}
\end{equation}
where $\mathcal{C}$ stands for $N \times N$ real matrix. In the limit of 
$N \to \infty$ the eigenvalue spectrum of a GinOE matrix is homogeneous 
and assumes a regular elliptic shape in the complex plane~\cite{sommers88}
\begin{displaymath}
p(\lambda) = \left\{ \begin{array}{cc}
 (\pi a b)^{-1}\ , & ({{\rm Re} z \over a})^2 + ({{\rm Im} z 
\over b})^2 \le 1\\
 0\ , & ({{\rm Re} z \over a})^2 + ({{\rm Im} z \over b})^2 > 1, 
  \end{array} \right.
\label{ginoelimit}
\end{displaymath}
where $a=1+\gamma,\ b=1-\gamma$ and $\gamma$ parametrizes a degree of 
matrix symmetry ($\gamma=1,\ \gamma=-1$ correspond to, respectively, 
symmetric matrix with all eigenvalues being real and antisymmetric matrix 
with imaginary eigenvalues, while $\gamma=0$ means full asymmetry). In 
physical situations with finite $N$, these spectra, however, loose their 
homegenity due to excess of real eigenvalues $\lambda_{\rm Re}$ whose 
expected number expressed as a fraction of $N$ in the $N \to \infty$ limit 
reads~\cite{edelman94}
\begin{equation}
\lim_{N \rightarrow \infty} {E_{\lambda_{\rm Re}} (N) \over N} = \sqrt{2 
/ (N \pi)}\ .
\label{ginoeexpect}
\end{equation}
A typical eigenvalue p.d.f. in the complex plane of a random matrix ($N = 
30$) obtained from 100000 independent matrix realizations is displayed in 
Figure 1.

\begin{figure}
\epsfxsize 6.5cm
\hspace{-0.5cm}
\epsffile{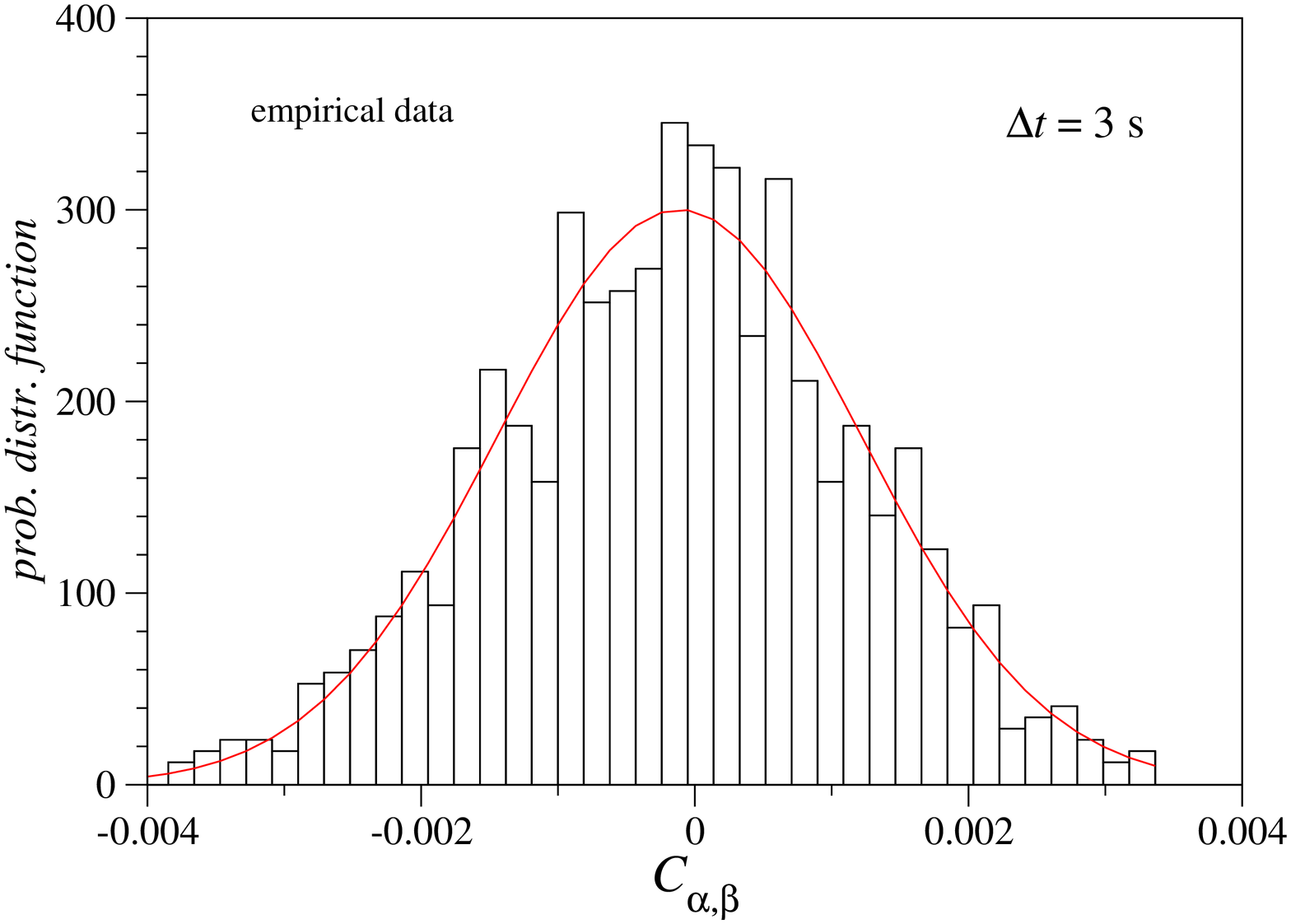}
\hspace{0.0cm}
\epsfxsize 6.95cm
\epsffile{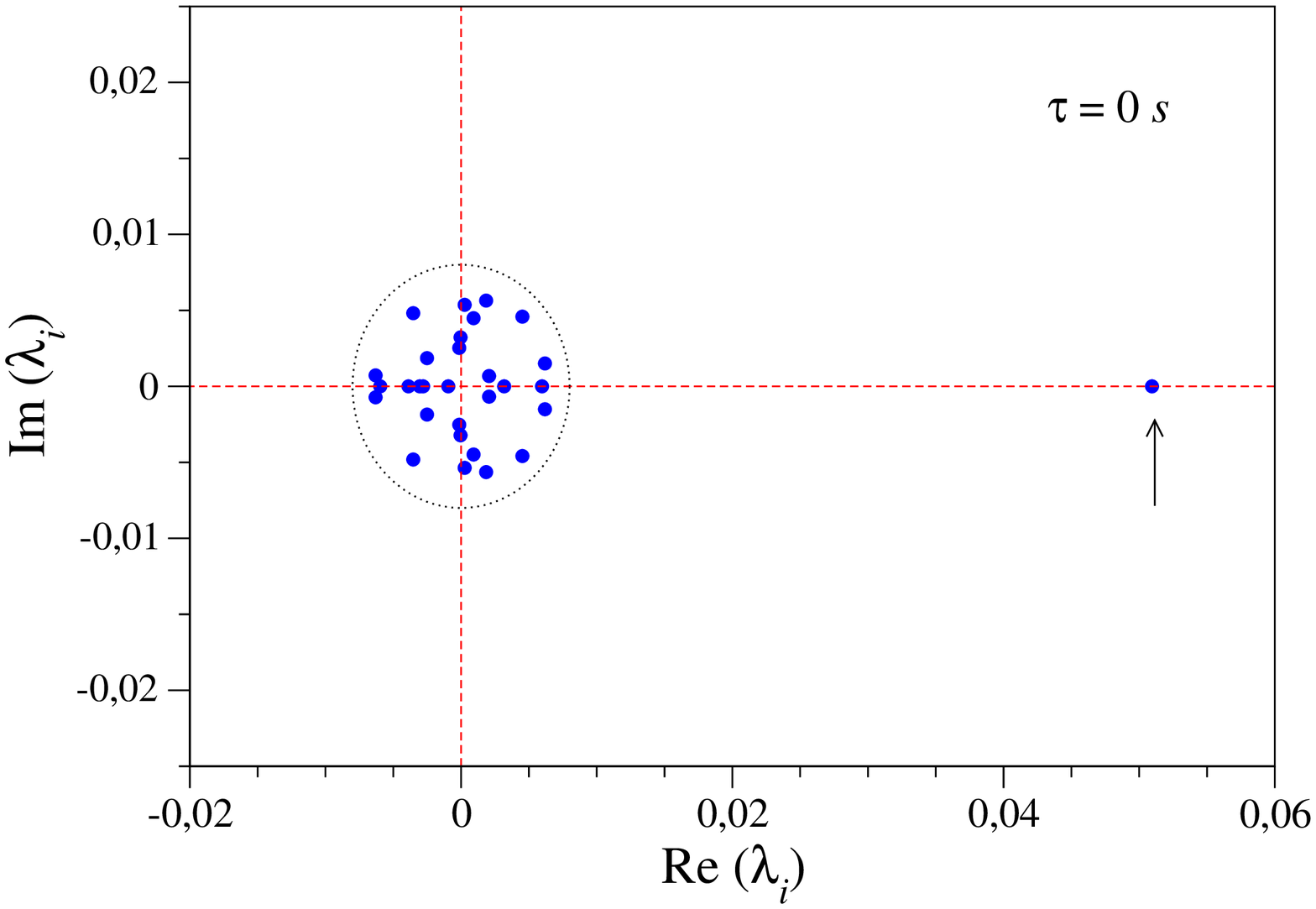}
\caption{(a) Probability density function of empirical correlation matrix 
(histogram) together with fitted Gaussian distribution (red solid line) 
for $\Delta t=3$ s and for zero time lag. (b) Spectrum of complex 
eigenvalues of correlation matrix for the same data as in (a). The largest 
real eigenvalue is pointed by an arrow. Dashed circle denotes theoretical 
eigenvalue spectrum for GinOE multiplied by standard deviation of 
matrix elements.}
\end{figure}

\section{Results}

Our example of an application of the asymmetric correlation matrix is
based on high frequency data from NYSE and Deutsche B\"orse~\cite{data}
spanning the interval 1 Dec 1997 $-$ 31 Dec 1999. We analyze $N=30$ stocks
belonging to the Dow Jones Industrials group and the same number of stocks
constituting the main German DAX30 index. We calculate each element of
${\bf C}$ by cross-correlating the time series pairs representing all
possible combinations of an American and a German stock. We neither
consider the correlations inside the German market nor inside the American
one. In order to investigate temporal dependencies between both markets we
introduce a time lag $\tau$ and associate it with the German stocks, i.e.
we look at $\{x_i^{(s)}\}_{i=1,...,T}$ and
$\{y_{i+\tau}^{(t)}\}_{i=1,...,T}$, where $\tau$ can assume both positive
and negative integer values. Thus, $\tau>0$ denotes a retardation of all
the signals corresponding to German stocks while $\tau < 0$ denotes the
opposite case.

\begin{figure}
\epsfxsize 6.5cm
\hspace{-0.5cm}   
\epsffile{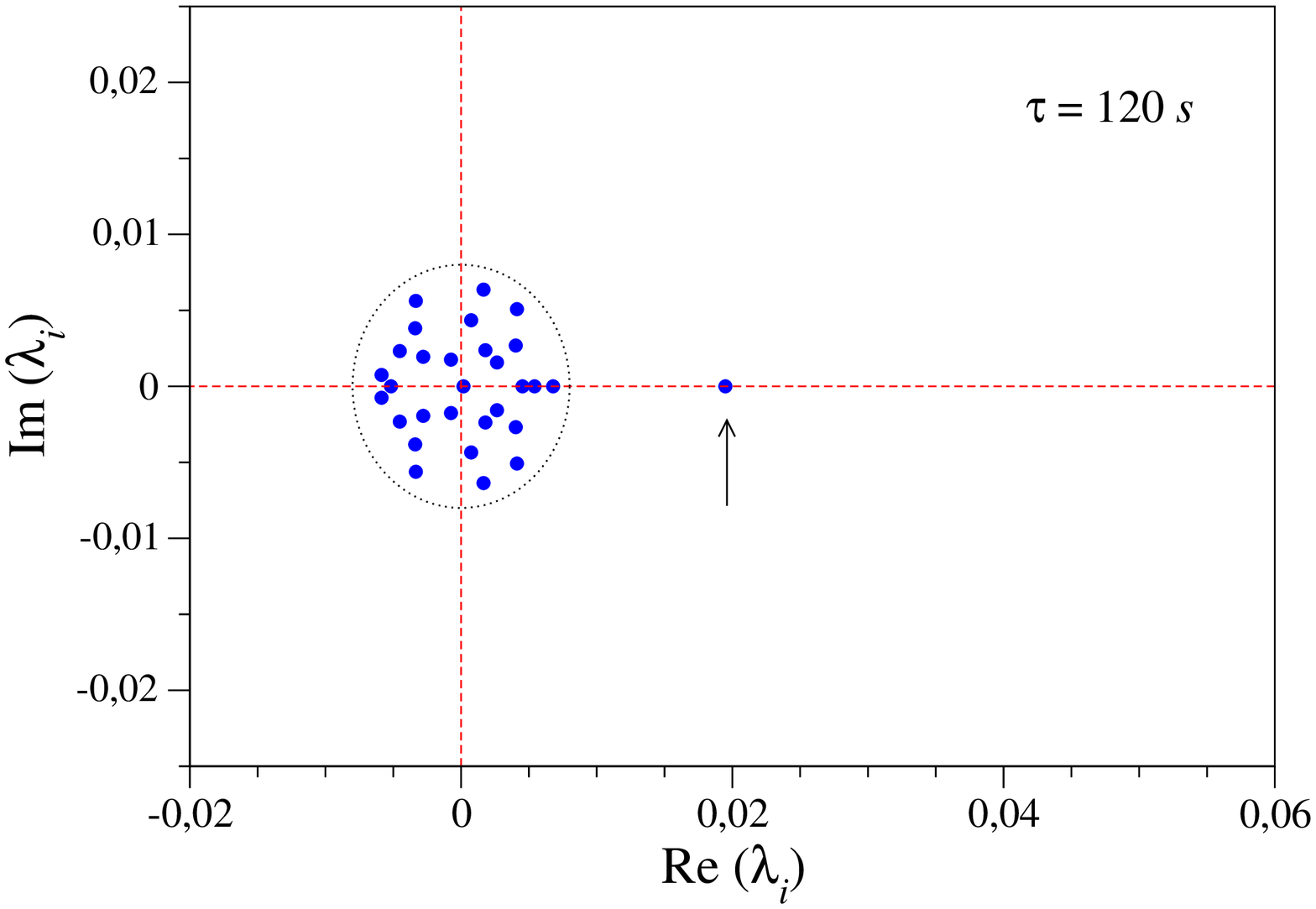}
\hspace{0.0cm}
\epsfxsize 6.5cm
\epsffile{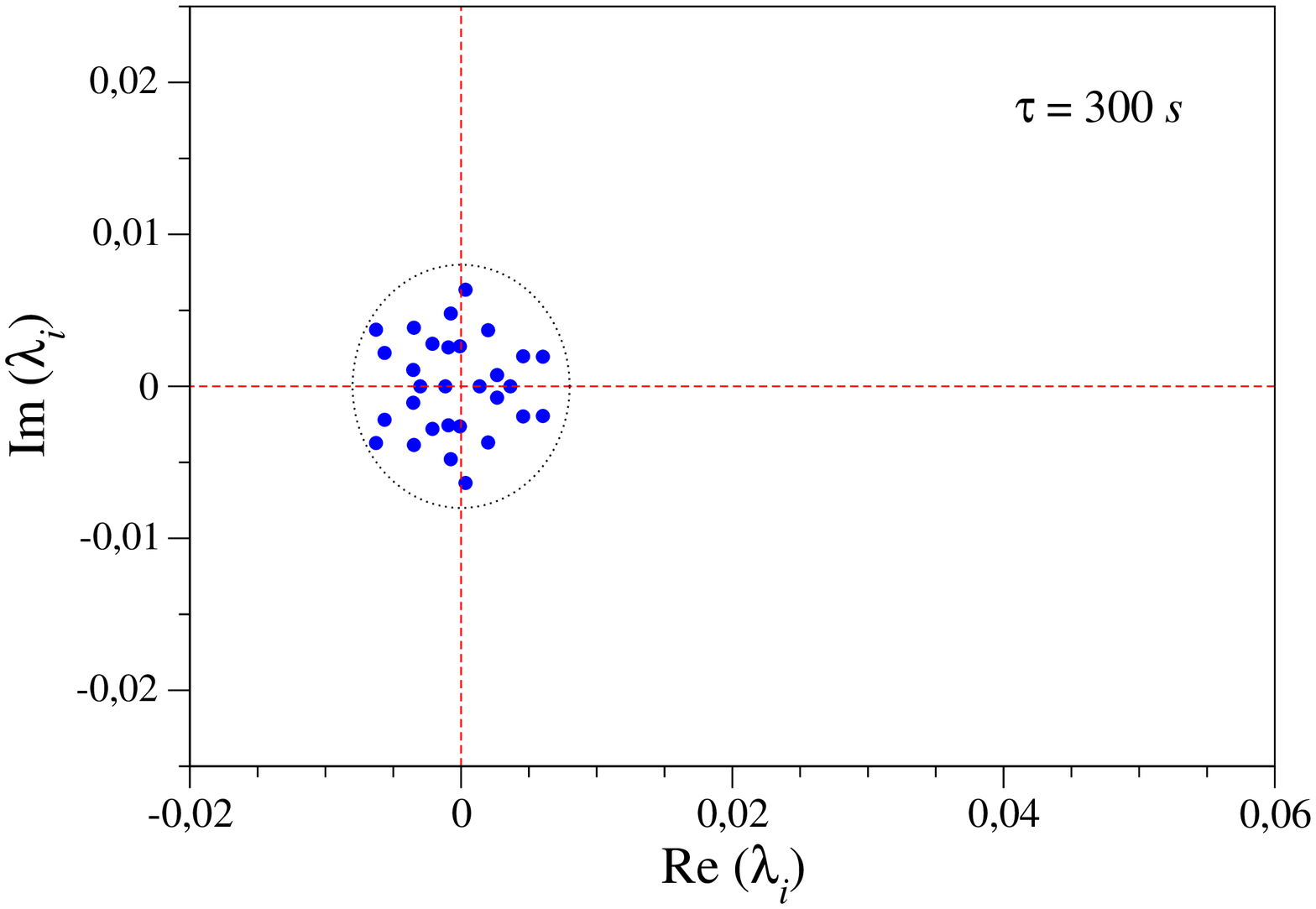}
\caption{Eigenvalue spectra in complex plane for empirical correlation 
matrix calculated for different values of time lag: $\tau = 120$ s (a) 
and $\tau=300$ s (b).}

\end{figure}

Since the two markets under study are separated by a few time zones, their 
activities overlap only for a relatively short period of a trading day 
(only the days that were common to both markets are considered). For most 
time it was only 90 minutes a day from 9:30 to 11:00 New York time (15:30 
to 17:00 Frankfurt time) and only after changing the trading hours in the 
German floor starting from 20 Sep 1999 the overlap interval increased to 
120 minutes (from 9:30 to 11:30 in New York and 15:30 to 17:30 in 
Frankfurt). This means that actually we can analyze the time series 
spanning 47700 minutes total. A good time resolution should be a crucial 
aspect of our analysis hence we consider only short time scales of 
returns: from $\Delta t=120$ seconds down to $\Delta t=3$ seconds. Shorter 
time scales cannot be used due to a fact that transaction times in the TAQ 
database are stored with only 1 s resolution.

\begin{figure}
\epsfxsize 6.5cm   
\hspace{-0.5cm}
\epsffile{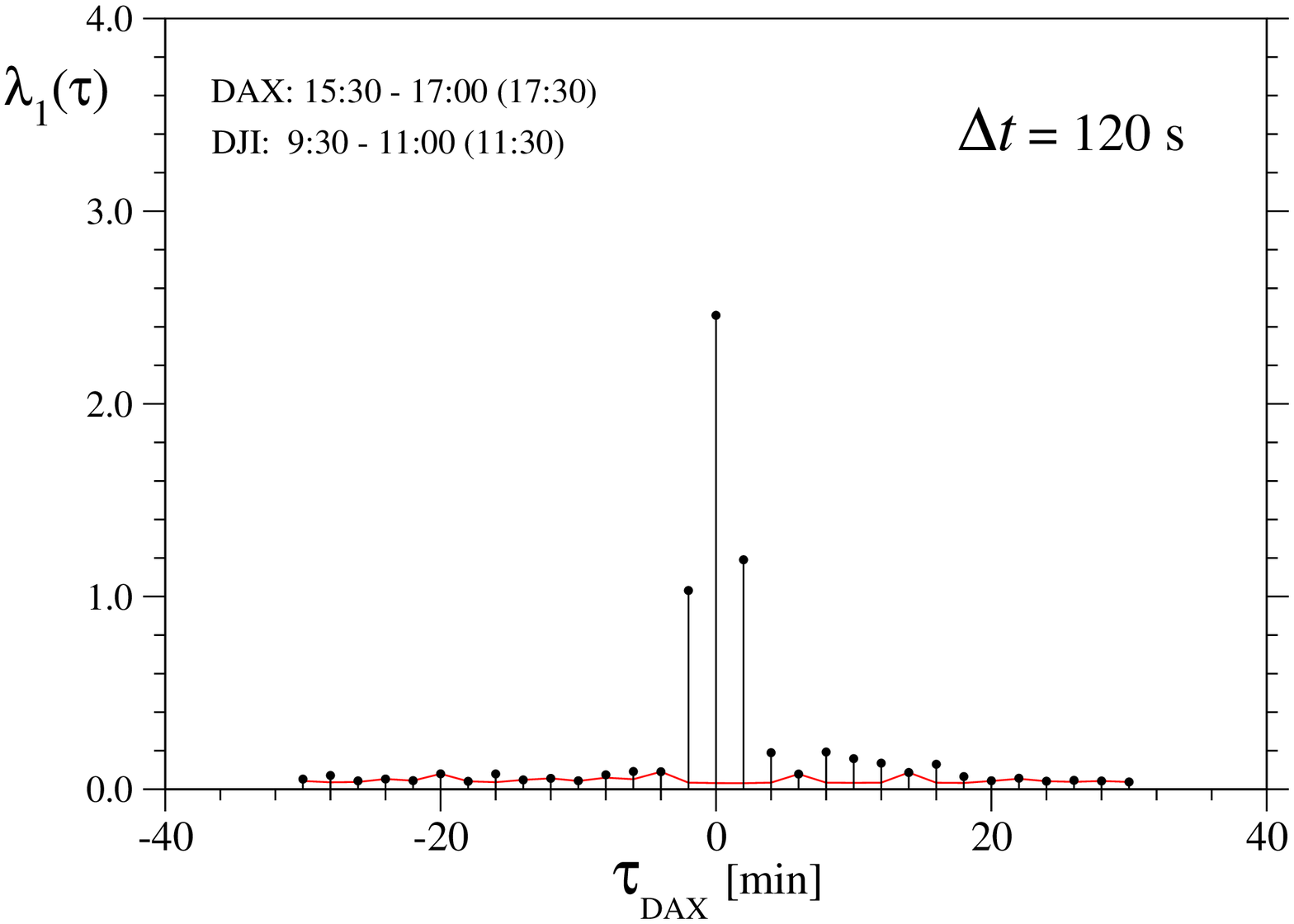}
\hspace{0.0cm}
\epsfxsize 6.5cm
\epsffile{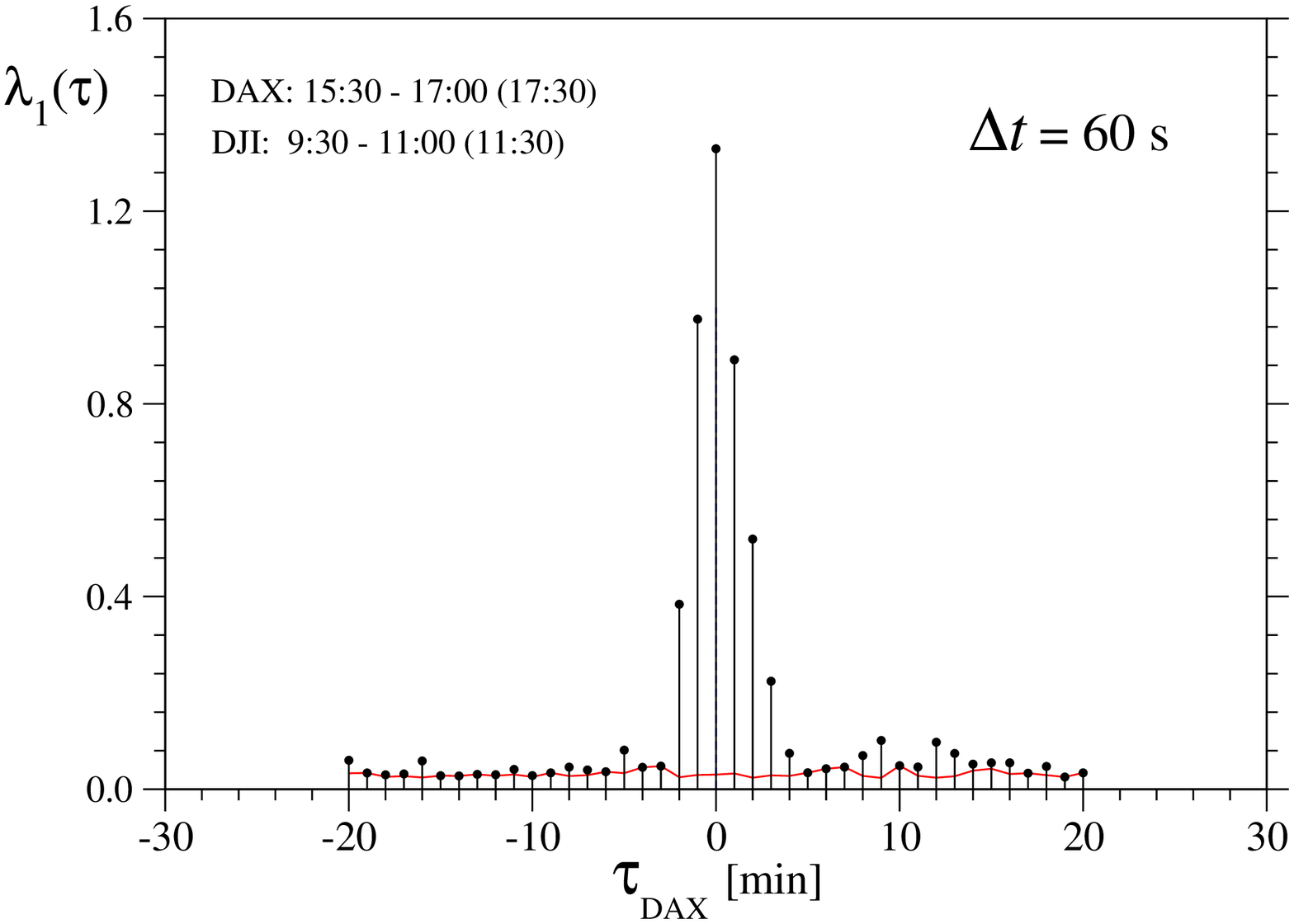}

\epsfxsize 6.5cm
\hspace{-0.5cm}
\epsffile{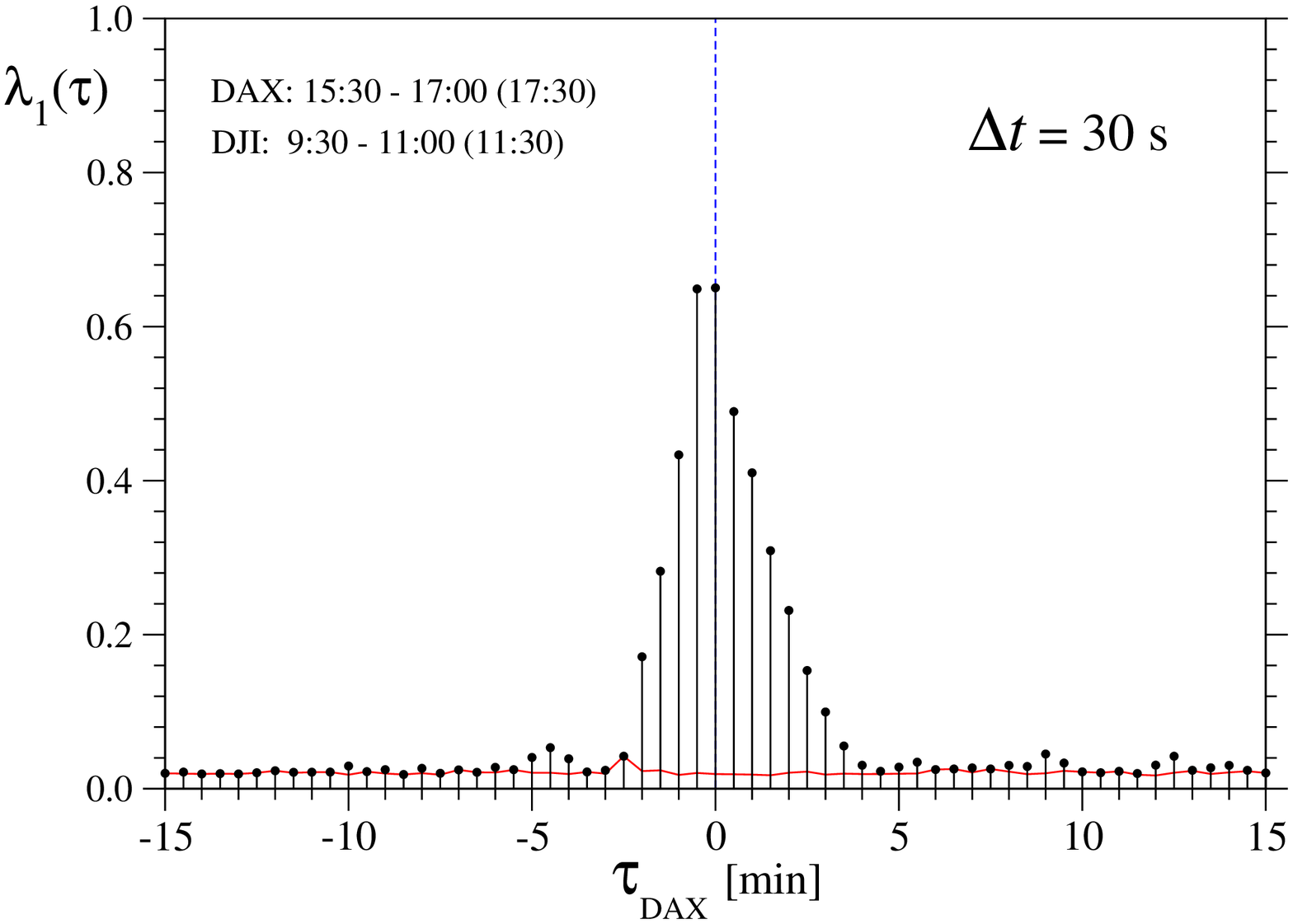}
\hspace{0.0cm}
\epsfxsize 6.5cm
\epsffile{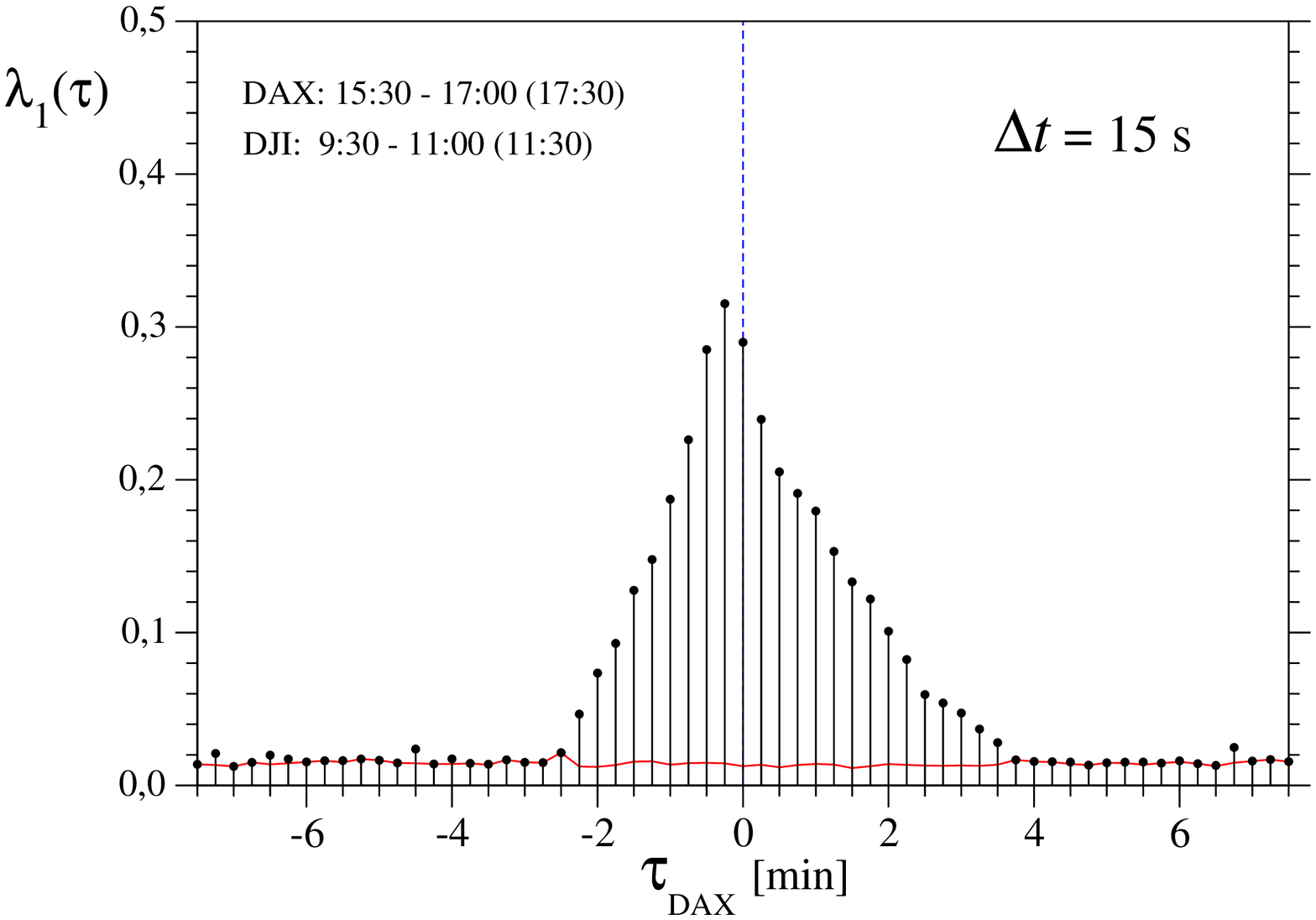}
\caption{$|\lambda_1(\tau)|$ (vertical lines and full circles) and 
$|\lambda_2(\tau)|$ (red solid line) for a few different time scales of 
returns: $\Delta t=120$ s (a), $\Delta t=60$ s (b), $\Delta t=30$ s (c) 
and $\Delta t=15$ s (d).}
\end{figure}

First of all let us look at the correlation matrix and its eigenvalues for
$\tau=0$ (no time lag, synchronous evolution of both markets). Figure 2(a)
presents p.d.f. of the matrix elements $C_{s,t}$ for $\Delta t=3$ s
(histogram) together with a fitted Gaussian distribution (red solid line).
Except the central part of the empirical distribution, where there are
excessive small positive elements and lacking small negative ones, the
Gaussian is well approximated by the histogram (the same refers to the
other time scales). Thus, the correlation matrix can be
treated~\cite{drozdz00} as a sum of an essentially random core matrix and
a non-random part carrying the actual inter-market correlations. This
suggests that we can expect the eigenvalue spectrum consisting of an RMT
bulk and at least one significant non-random eigenvalue responsible for
the correlations. In fact, exactly this type of spectrum can be seen in
Figure 2(b). All except one eigenvalues are localized inside the RMT
prediction for a completely asymmetric matrix and the remaining largest
one is distant and resides on the real axis. By an analogy to a
symmetric matrix we are justified to associate this eigenvalue with the
coupling strength of the two markets (the global market factor).
Interestingly, there is no other factor which can influence the behaviour
of some smaller parts of the markets like e.g. specific economic sectors.

\begin{figure}   
\epsfxsize 10cm
\hspace{1.0cm}
\epsffile{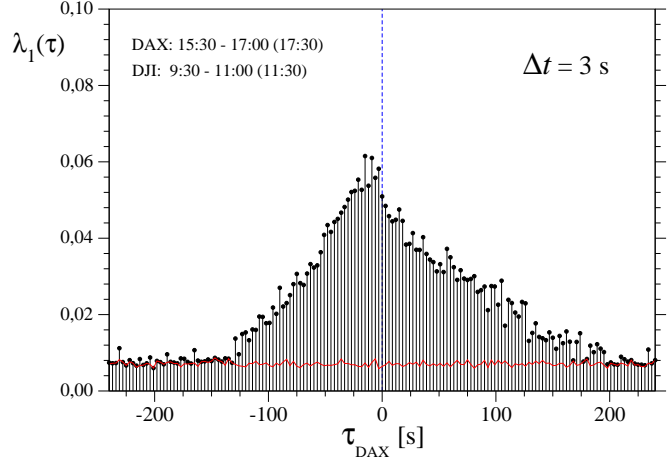}
\caption{$|\lambda_1(\tau)|$ (vertical lines and full circles) and
$|\lambda_2(\tau)|$ (red solid line) for the shortest time scale $\Delta 
t=3$ s.}
\end{figure}

Figure 3 shows examples of the eigenspectra for two different positive 
time lags. As we increase $\tau$ from 0 s up to 5 min, we observe a 
gradual decrease of $|\lambda_1|$ which remains real even for $\tau>120$ 
s, but eventually looses its identity by drowning in the sea of random 
eigenvalues for $\Delta t=5$ min. From the market perspective, after such 
a time interval the stocks traded in Frankfurt forget about what happened 
earlier in New York. We however still cannot say anything decisive about 
the possible directional information flow between the markets. It requires 
a more systematic investigation in which the largest eigenvalue 
$\lambda_1$ (i.e. the one with the largest absolute magnitude) becomes a 
function of variable $\tau$. Figure 4 displays $\lambda_1(\tau)$ for 
different time scales of the returns. It can be seen that with the 
resolutions of $\Delta t=120$ and 60 s the maximum coupling between the 
markets occurs for synchronous signals and the non-random correlations 
exist for $-3 \le \tau \le 3$ minutes. For $\Delta t=30$ s a weak trace of 
asymmetry in both the maximum position and the memory length can be 
identified, which is confirmed in the plot for $\Delta t=15$ s. Going down 
to the shortest time scale of 3 s, this asymmetry becomes clear. Figure 5 
documents that the stocks from both markets are maximally correlated if 
the American market is retarded by about 3-15 seconds in respect to its 
German counterpart.  This observation is somehow counterintuitive because 
one might expect that the American stock market, being the largest in the 
world and representing the world's largest economy, is less dependent on 
external influence than is the German market. We cannot give a 
straightforward explanation of this phenomenon, though. Its source can lie 
in memory properties of the American market as well as in some specific 
behaviour of investors in the beginning of a trading day in New York. For 
example, they may carefully observe the evolution of the European markets 
which in the years 1998-99 used to finish their activity rather soon after 
the American markets had been opened. We also cannot exclude the 
possibility that the reason for this is a possible existence of artifacts 
in the trade recordings in TAQ or KKMDB databases which cannot be 
identified in data. Finally, the observed asymmetry of the curve tails in 
Figures 4 and 5 with respect to $\tau=0$ can be explained, at least in 
part, by different autocorrelation properties of the two markets under 
study. This is evident in Figure 6, where the largest eigenvalue of the 
symmetric matrix ${\bf C}_{\rm XX}$ is calculated separately for the 
German and for the American markets. Here $\tau$ assumes only non-negative 
values due to a symmetry of the problem; $\lambda_1(\tau)$ is a 
multivariate counterpart of the autocorrelation function. It is clear from 
Figure 6 that the German market has considerably longer and stronger 
memory than its American counterpart;  in fact, this memory can lead to 
longer-lasting cross-dependencies presented in Figures 4 and 5 if the 
German market is retarded. On the other hand, investors in Frankfurt may 
need a longer time to collect all the information needed before they make 
investment decisions if they take more markets and more information into 
consideration. It is also possible that the American stock market is 
technically more advanced and, on average, allows the investors to react 
quicker than in Germany.

\begin{figure}
\epsfxsize 10cm
\hspace{1.0cm}
\epsffile{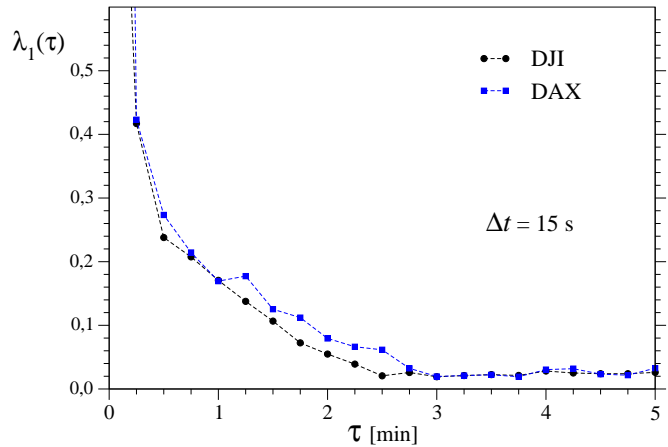}
\caption{$|\lambda_1(\tau)|$ for symmetric correlation matrix 
${\bf C}_{\rm XX}$ calculated for the American (black circles) and the 
German (blue squares) stocks separately. Longer and stronger memory in the 
latter case is visible.}
\end{figure}

\section{Conclusions}

We construct an asymmetric real correlation matrix from time series of 
returns representing two separate groups of stocks: German and American ones.
Nonexistence of a symmetry condition allows us to concentrate solely on 
the inter-market correlations without mixing them with the correlations 
that are inner to only one market, and to study temporal properties of 
such correlations. We introduce a time lag associated with German stocks 
and investigate traces of direct information transfer from one market to 
the other which can manifest itself in the existence of significant 
non-synchronous couplings between the markets represented by a 
$\tau$-shifted maximum in the largest eigenvalue of the empirical 
correlation matrix. We identified such delayed correlations indicating 
that the same information is shared by both markets with the American one 
following its German counterpart only after a few seconds. This 
observation, however, cannot be treated as a fully convincing one due to a 
significant broadening of the $\lambda_1(\tau)$ maximum and an 
unintuitive direction of this transfer from a smaller towards a larger 
market. Another conclusion from our results is that the coupling between 
the two analyzed markets is only of a one-factor type. We do not noticed 
other, more subtle partial couplings that can involve a subset of stocks.

Our results can be compared with the results of ref.~\cite{toth06} in 
which an analysis of the delayed correlations between different stocks 
traded on the American market are studied by means of the correlation 
coefficients. It is worth mentioning that a similar analysis can also be 
performed by applying the asymmetric correlation matrices used in our 
work.


\begin{thebibliography}{99}

\bibitem{mantegna99} R.N.~Mantegna, Eur.~Phys.~J.~B {\bf 11}, 193-197 
(1999)

\bibitem{bonanno00} G.~Bonanno, N.~Vandewalle, R.N.~Mantegna, Phys.~Rev.~E 
{\bf 62}, R7615-R7618 (2000)

\bibitem{giada01} L.~Giada, M.~Marsili, Phys.~Rev.~E {\bf63}, 061101
(2001)

\bibitem{plerou02} V.~Plerou, P.~Gopikrishnan, B.~Rosenow, L.A.N.~Amaral,
T.~Guhr, H.E.~Stanley, Phys.~Rev.~E {\bf 65}, 066126 (2002)

\bibitem{onnela03} J.-P.~Onnela, A.~Chakraborti, K.~Kaski, J.~Kert\'esz, 
A.~Kanto, Phys.~Rev.~E {\bf 68}, 056110 (2003)

\bibitem{dimatteo04} T.~Di Matteo, T.~Aste, R.N.~Mantegna, Physica A {\bf 
339}, 181-188 (2004)

\bibitem{kim05} D.-H.~Kim, H.~Jeong, Phys.~Rev.~E {\bf 72}, 046133 (2005)

\bibitem{mizuno05} T.~Mizuno, H.~Takayasu, M.~Takayasu, preprint
physics/0508164 (2005)

\bibitem{roll80} R.~Roll, S.A.~Ross, J.~Finance {\bf 35}, 121-130 (1980)

\bibitem{noh00} J.-D.~Noh, Phys.~Rev.~E {\bf 61}, 5981-5982 (2000)

\bibitem{laloux99} L.~Laloux, P.~Cizeau, J.-P.~Bouchaud, M.~Potters,
Phys.~Rev.~Lett. {\bf 83}, 1467-1470 (1999); V.~Plerou, P.~Gopikrishnan,
B.~Rosenow, L.A.N.~Amaral, H.E.~Stanley, Phys.~Rev.~Lett. {\bf 83},
1471-1474 (1999)

\bibitem{kwapien06} J.~Kwapie\'n, S.~Dro\.zd\.z, P.~O\'swi\c ecimka,
Physica A {\bf 359}, 589-606 (2006)

\bibitem{bonanno01} G.~Bonanno, F.~Lillo, R.N.~Mantegna, Quant.~Finance 
{\bf 1}, 96-104 (2001)

\bibitem{gorski06} A.Z.~G\'orski, S.~Dro\.zd\.z, J.~Kwapie\'n, 
P.~O\'swi\c ecimka, Acta.~Phys.~Pol.~B, this issue (2006)

\bibitem{ginibre65} J.~Ginibre, J.~Math.~Phys.~{\bf 6}, 440-449 (1965)

\bibitem{timme04} M.~Timme, F.~Wolf, T.~Geisel, Phys.~Rev.~Lett.~{\bf 
92}, 074101 (2004)

\bibitem{fyodorov97} Y.~Fyodorov, B.~Khoruzhenko, H.-J.~Sommers, 
Phys.~Rev.~Lett.~{\bf 79}, 557-560 (1997)

\bibitem{stephanov96} M.~Stephanov, Phys.~Rev.~Lett.~{\bf 76}, 4472 (1996)

\bibitem{akemann04} G.~Akemann, T.~Wettig, Phys.~Rev.~Lett.~{\bf 92}, 
102002 (2004)

\bibitem{kwapien00} J.~Kwapie\'n, S.~Dro\.zd\.z, A.A.~Ioannides, 
Phys.~Rev.~E {\bf 62}, 5557-5564 (2000)

\bibitem{sommers88} H.-J.~Sommers, A.~Crisanti, H.~Sompolinsky, Y.~Stein, 
Phys.~Rev.~Lett.~{\bf 60}, 1895-1898 (1988)

\bibitem{edelman94} A.~Edelman, E.~Kostlan, M.~Shub, 
J.~Am.~Math.~Soc.~{\bf 7}, 247-267 (1994)

\bibitem{kanzieper05} E.~Kanzieper, G.~Akemann, Phys.~Rev.~Lett.~{\bf 95}, 
230201 (2005)

\bibitem{data} http://www.taq.com (data from NYSE) and H. Goeppl,
Karlsruher Kapitalmarktdatenbank (KKMDB), Institut f\"ur
Entscheidungstheorie u. Unternehmensforschung, Universit\"at Karlsruhe
(TH) (data from Deutsche B\"orse)

\bibitem{drozdz00} S.~Dro\.zd\.z, A.Z.~G\'orski, F.~Ruf, J.~Speth, Physica 
A {\bf 287}, 440-449 (2000) 

\bibitem{toth06} B.~T\'oth, J.~Kert\'esz, Physica A {\bf 360}, 505-515 
(2006)

\end{thebibliography}
\end{document}